 \documentstyle[aaspp]{article}

\def\etal{{\it et~al.}}
\def\ie{{\it i.e.,\ }}
\def\eg{{\it e.g.,\ }}
\def\cf{{\it cf.,\ }}

\def\vfp{Paper~I}
\def\HDF{HDF}
\def\HST{HST}
\def\DEEP{DEEP}
\def\LRIS{LRIS}

\def\m{m}

\def\kpc-1{{kpc$^{-1}$}}
\def\Mpc-1{{Mpc$^{-1}$}}
\def\s-1{{sec$^{-1}$}}
\def\pdeg2{{deg$^{-2}$}}
\def\deg{$^{\circ}$}
\def\h0{{$H_0$}}
\def\q0{{$q_0$}}
\def\z{{$z$}}

\def\Hconst#1{{$H_0 = $ {#1} {\rm km~s$^{-1}$~Mpc$^{-1}$}}}
\def\wave#1{{\rm ~$\lambda${#1}}}
\def\Halpha{{H$\alpha$}}
\def\Hbeta{{H$\beta$}}

\def\HI   {{{\sc H\thinspace i}}}
\def\fNII {{{\sc [N\thinspace ii]}}}
\def\fOII {{{\sc [O\thinspace ii]}}}
\def\fOIII{{{\sc [O\thinspace iii]}}}
\def\I{{$I_{814}$}}
\def\V{{$V_{606}$}}
\def\Vmax{{$V_{term}$}}
\def\SNr{S/N}

\slugcomment{Accepted by {\it The Astrophysical Journal Letters}}

\begin{document}

\title{Optical Rotation Curves of Distant Field Galaxies: \\ 
Sub-$L^\star$ Systems$^{1,2}$}

\author{Nicole P. Vogt, Andrew C. Phillips, 
S. M. Faber, Jes\'us Gallego$^3$, \\ Caryl Gronwall, R. Guzm\'an, 
Garth D. Illingworth, David C. Koo, \& J. D. Lowenthal$^4$}

\affil{University of California Observatories / Lick Observatory, 
Board of Studies in Astronomy and Astrophysics, 
University of California, Santa Cruz, CA 95064}

\altaffiltext{1}{Based on observations obtained at the W. M. Keck Observatory,
which is operated jointly by the California Institute of Technology and the
University of California.}

\altaffiltext{2}{Based in part on observations with the NASA/ESA {\it Hubble
Space Telescope}, obtained at the Space Telescope Science Institute, which is
operated by AURA, Inc., under NASA contract NAS 5--26555.}

\altaffiltext{3}{Current address: Dpto de Astrofisica Universidad Complutense,
Madrid, E-28040 Spain}

\altaffiltext{4}{Hubble Fellow}

\begin{abstract}

Moderate-resolution spectroscopic observations from the Keck 10\m\ telescope
are used to derive internal kinematics for eight faint disk galaxies in the
fields flanking the Hubble Deep Field.  The spectroscopic data are combined
with high--resolution $F814W$ WFPC2 images from the {\it Hubble Space
Telescope} which provide morphologies and scale-lengths, inclinations and
orientations.  The eight galaxies have redshifts $0.15 \lesssim z \lesssim
0.75$, magnitudes $18.6 \leq I_{814} \leq 22.1$ and luminosities $-21.8 \leq
M_B \leq -19.0$ (\Hconst{75} and $q_0 = 0.05$).  Terminal disk velocities are
derived from the spatially-resolved velocity profiles by modeling the effects
of seeing, slit width, slit misalignment with galaxy major axis, and
inclination for each source.  These data are combined with the sample of Vogt
\etal\ (1996) to provide a high-redshift Tully-Fisher relation that spans three
magnitudes.  This sample was selected primarily by morphology and magnitude,
rather than color or spectral features.  We find no obvious change in the shape
or slope of the relation with respect to the local Tully-Fisher relation.  The
small offset of $\lesssim 0.4$ $B$ mag with respect to the local relation is
presumably caused by luminosity evolution in the field galaxy population, and
does not correlate with galaxy mass.  A comparison of disk surface brightness
between local and high-redshift samples yields a similar offset, $\sim$0.6 mag.
These results provide further evidence for only a {\it modest} increase in
luminosity with lookback time.

\end{abstract}


\section{Introduction}

Until recently, studies of faint field galaxies have been limited to galaxy
counts, colors, and redshift distributions, which can be used to construct
luminosity functions at earlier epochs. Such luminosity functions are then
compared to models incorporating a certain amount of number evolution
(controlled by galaxy formation and merging) coupled with luminosity and color
evolution (controlled by star formation histories).  Recent models range from
those predicting only a small degree of luminosity evolution (\eg Gronwall \&
Koo 1995) to those invoking entirely new classes of galaxies (\eg Babul \& Rees
1992; Babul \& Ferguson 1996), to those requiring high merger rates (\eg
Broadhurst, Ellis, \& Glazebrook 1992).  A direct measure of luminosity
evolution in field galaxies will help to distinguish between various
hypotheses.  Such measures have recently been attempted, but the results are
somewhat contradictory.  Schade \etal\ (1996a) found evidence for disk
brightening by 1.2 $B$ mag in galaxies at redshifts $0.5 \leq z \leq 1.2$. 
Simard \& Pritchet (1996) found even greater levels of evolution ($2.5 \pm 0.5$
mag) in a sample of very blue galaxies at $z \sim 0.35$ for which strong \fOII\
lines could be spatially resolved.  Rix \etal\ (1997), also using kinematic
information, derived a brightening of 1.5 mag at $z \sim 0.25$ for
sub-$L^\star$ galaxies.  In contrast, Vogt \etal\ (1996) and Bershady (1997),
using optical rotation curves, and Forbes \etal\ (1996), using line widths,
found only small deviations from the local Tully-Fisher (TF) relation for
spiral galaxies, implying only modest brightening ($\sim$0.4 mag) out to $z\sim
1$.  Simard \& Pritchet postulate that these various results could be
reconciled if strong luminosity evolution were present {\it only} in lower-mass
systems.  If confirmed, this would be an important factor in understanding the
evolution of field galaxies.

This Letter introduces well-resolved rotation curves of eight predominantly
lower-luminosity galaxies ($L_B \lesssim L_B^\star$ = -20.3, see Efstathiou,
Ellis, \& Peterson 1988), selected by morphology as suitable TF candidates. 
These new observations provide a valuable test of the mass-dependent luminosity
evolution hypothesis, particularly in comparison to the higher-mass sample
presented by Vogt \etal\ (1996; hereafter ``\vfp'').  Combined with the work of
\vfp, these data form a sample of rotation curves for 16 galaxies at redshifts
{0.15~$\lesssim$~\z~$\lesssim$~1}, ranging over half the age of the universe
(for $q_0 = 0$, one--third for $q_0 = 0.5$).  We also use this combined data
set to explore trends in surface brightness for comparison with Colless \etal\
(1994),  Forbes \etal\ (1996) and Schade \etal\ (1996a).  Detailed analysis of
a significantly larger data set is currently underway, and these results will
be used to explore a variety of relevant selection effects.  A full description
of our analysis techniques is deferred to that paper (Vogt \etal\ 1997).

\section{Observations}

The TF candidate objects were selected from WFPC2 F814W (``$I_{814}$'') images
of the flanking fields of the Hubble Deep Field (\HDF; Williams \etal\ 1996).
Selection was based on the following criteria:
({\it i}) undistorted disk morphology;
({\it ii}) inclination greater than 30\deg;
({\it iii}) no interacting companions or obscuring foreground stars; and
({\it iv}) $I_{814} \leq 22.5$.
The selection was made with no apriori knowledge of redshifts or luminosities. 
The galaxies were observed in 1996 April under the auspices of the \DEEP\
project (Koo 1995), using the Low Resolution Imaging Spectrograph (\LRIS;  Oke
\etal~1995) on the Keck 10\m\ telescope. \LRIS\ employs slitmasks to provide
``long-slit'' spectral observations for multiple objects simultaneously.  For
the TF candidates, the slitlet for each object was tilted to align with the
major axis of each galaxy.  Slitlets were {1\farcs 1} wide and $\geq${12\farcs
0} long for these objects.  Integration times were 50 minutes for each of two
600 line mm$^{-1}$ gratings, blazed at 5000 and 7500~\AA; the combined spectral
range was roughly 3800--8600~\AA.  Spectral and spatial scales were
1.28~\AA~pix$^{-1}$ and {0\farcs 215}~pix$^{-1}$, respectively.  The seeing was
approximately {1\farcs 0} FWHM. In additional to spectroscopy, two 300s
$V$-band images of the field were acquired.  (See Phillips \etal~1997 for more
details of the observations and for galaxy coordinates.)

The new spectral data have several significant improvements over those of \vfp.
The targets were preselected as suitable TF galaxies, whereas the rotation
curves in \vfp\ were obtained serendipitously.  Slitlets were aligned with the
galaxy major axes, removing a source of potentially significant error. 
Finally, the expanded spectral range means that multiple emission lines were
observed for each object, \eg \fOII\wave{3727} through \fOIII\wave{5007} for $z
< 0.7$ (the majority of our sources), and through \Halpha\ for $z < 0.3$.  
$z > 0.7$, only the \fOII\ lines were available.

\section{Data Reduction and Analysis}

\subsection{Spectral Measurements}

The \LRIS\ spectra were debiased and flat--fielded, and then rectified. 
Wavelength calibration was done using the procedure described in Kelson \etal\
1997.  No relative or absolute flux calibration was applied.  All spectra (not
just those preselected for TF work) were examined for spatially extended
emission lines, but only one additional object (IE4-1304-1007) displayed such
lines (the apparent inclination of this object was too low for our criteria,
but we include it in our final sample).  Among the 18 preselected candidates,
one has yielded no redshift identification; one has a pure early-type spectrum
with no detected emission lines; one at $z = 0.109$ displayed unusually weak
lines, both in emission and absorption; eight show normal disk galaxy spectra
but the emission lines are too weak to use to derive velocity curves; and seven
display emission lines of sufficient strength to determine velocity structure. 
The group of eight has virtually the same median redshift (0.50) and redshift
range (0.41--0.77) as do the seven successful targets.

Both the \fOII\wave{3727} doublet and \Hbeta\ and \fOIII\wave{5007} were
observed for most sources in the final sample.  The spatially-resolved emission
lines were analyzed using the same Gaussian profile fitting technique described
in \vfp\ (see also Vogt 1995).  Briefly, a single (or double) Gaussian profile
was fit to each emission line (or doublet) at each point in the spatial
direction.  Profiles were considered acceptable whenever the Gaussian fit met
minimum requirements in height and width, generally a signal-to-noise-ratio
(\SNr) of 5$\sigma$ and 3$\sigma$, respectively;  the typical value was
10$\sigma$ for both width and amplitude.  Central wavelengths of the profiles
were used to construct observed rotation curves.

As discussed in \vfp, the sizes of the disks in these galaxies are on the order
of the seeing and the slit--width.  Thus, wavelength shifts in the observed
spectral lines are not only a function of the velocity profile of the disk but
also the surface brightness distribution and the mapping of the
seeing--smoothed flux through the full slit. (A simplifying factor for the new
data is that the misalignment of the slit relative to the galaxy major axis was
less than 10\deg\ for all but one source.) To derive terminal velocities we
must correct for these effects. To this end, we employ a grid of simple
exponential disk models with different terminal velocities to simulate emission
lines, and fit these model emission lines identically to the spectral data. 
The circular velocity of the galaxy model was iteratively adjusted to match the
observed data, and adopted as the intrinsic terminal velocity, \Vmax.  (For
some galaxies, a clear turnover in the velocity curve was not observed.  Rather
than match the inner, rising rotation curve and extrapolate to a turnover
velocity, we have chosen the model whose measured turnover velocity matches the
maximum velocity measured in our spectra; these models provide lower limits to
the true \Vmax.) Errors in \Vmax\ were estimated by varying the inclination and
position angle of each galaxy by $\pm$~10\deg.

\vfill\eject
\subsection{Photometric Parameters}

The \HST\ \I\ images were analyzed using IRAF-based tools (see \eg Forbes
\etal\ 1994; Phillips \etal\ 1997).  Total magnitudes were measured from
aperture growth curves; inclinations and position angles were estimated from
outer elliptical isophotes; and disk scale lengths were measured from
simultaneous disk-plus-bulge fits to the major axis intensity profiles.  \HST\
took images of the flanking fields in \I\ only, so a \LRIS\ $V$ image was used
to determine a $V$--$I$ color.  The \I\ image was seeing-degraded to match the
ground--based image, and the color was determined within a {2\farcs 2} diameter
aperture (see Phillips \etal\ 1997 for more detail).

Intrinsic galaxy parameters were calculated using the measured redshifts,
photometry, and angular scales, assuming \Hconst{75} and $q_0 = 0.05$.$^5$ To
determine restframe $L_B$, $k$-corrections were interpolated from the model
SEDs of Gronwall \& Koo (1995), which are based on Bruzual \& Charlot (1993)
models and realistic star-formation scenarios.  Current epoch (\ie
non-evolving) SEDs were used.  Since restframe $B$ corresponds to observed \V\
at $z \sim 0.4$ and to \I\ at $z \sim 0.8$, errors in the $k$-corrections
should be small.  Galactic extinction was taken to be negligible for the \HDF\
(Williams \etal~1996), and sources were corrected for internal extinction
following the method of Tully \& Fouqu\'e (1985) in order to be consistent with
Pierce \& Tully (1988; 1992).  

\altaffiltext{5}{Data from \vfp\ have been adjusted for these new values.}

\section{Results and Discussion}

Images of the eight new galaxies and their spatially-resolved \fOII\ lines are
shown in Figure~\ref{PLATE} ({\sc Plates NN1--NN2)}), along with the observed
and modeled velocity curves.  Like the eight galaxies discussed in \vfp, these
new distant TF galaxies appear to be quite similar to local normal spiral
galaxies, both morphologically and kinematically.  The HST images show
apparently normal, disk-dominated spirals.  Allowing for seeing and resolution
effects, the velocity curves are qualitatively similar to those of local
spirals.  The rotation curves are traceable to $\sim$3 exponential scale
lengths ($R_d$) in the disks, a length comparable to the extent of rotation
curves for local galaxies (\cf Vogt 1995).  A simple estimate of their masses,
$M = V^2 R / G$, yields values of 0.5--$5 \times 10^{11} M_{\odot}$ within 15
kpc, similar to the range of masses found for nearby spirals.

For purposes of discussion, we separate the velocity data into two classes. 
High quality is defined as having sufficient \SNr\ and resolution elements to
clearly determine a terminal velocity; at least one emission line free of
strong night-sky contamination; apparent inclination greater than 30\deg; and
slit misaligned with the galaxy major axis by {$<$20\deg}.  Six of the eight
new galaxies and five from \vfp\ meet these criteria.  Note that the
serendipitous object, IE4-1304-1007, is a low-quality source.

\subsection{The High-Redshift Tully-Fisher Relation}

In Figure~\ref{TF} we compare the 16 galaxies to a local TF relation in the
restframe $B$-band. The local relation shown is an {\it inverse} fit (\ie
\Vmax\ as a function of $M_B$) to the local sample from Pierce \& Tully (1992);
this is the proper fit for comparison with a magnitude--limited sample (see
\vfp).  The 11 high-quality sources have a weighted offset of $0.36 \pm 0.13$
mag relative to the local relation, and an {\it rms} dispersion of 0.65 mag. 
This observed dispersion matches the combined estimated errors of the
logarithmic velocity widths (0.47), the restframe $B$ magnitudes (0.2), and an
assumed intrinsic scatter in the TF relation (0.4; \cf Willick \etal~1996 and
references therein), thus helping to validate our error estimates.  The
lower-quality points show a much larger scatter, as expected.

We emphasize that the derived offset, $\sim$0.4 mag, represents an {\it upper
limit} to luminosity evolution of field galaxies, for these reasons: any
magnitude-limited sample is biased towards more luminous objects; our analysis
is restricted to objects with detectable emission lines --- that is, actively
star-forming galaxies which are likely to have elevated $B$ luminosities; some
terminal velocities are lower limits; and we may be overcorrecting for
extinction if galaxies were less dusty at earlier epochs.  Our choice of $q_0 =
0.05$ is also conservative --- derived luminosities are reduced by 0.1--0.4 mag
for $q_0 = 0.5$.  While mass evolution is a possible factor, we have implicitly
assumed that evolution in luminosity dominates the observed offset.  We do not
expect the masses to evolve strongly, though, given the presence of clearly
formed disks.  

The new \HDF\ data extend the luminosity range of our total sample to $-21.8
\lesssim M_B \lesssim -19$. It is notable that {\it there is no deviation from
a linear relation over this range}, \ie there is no evidence in these data for
different amounts of luminosity evolution in different luminosity or mass
regimes.  

One explanation for the wide range in luminosity evolution found by various
groups is that sample selection strongly affects the degree of evolution
detected in a given sample.  In our sample, the galaxies were chosen primarily
by morphology.  On the other hand,  Bershady (1997), Simard \& Pritchet (1996),
and Rix \etal~(1997) all selected {\it blue} galaxies.  These studies all had
different sample selection, redshift ranges, and observational and analysis
techniques;  a direct comparison is not practical nor is a full discussion of
these parameters within the scope of this Letter (see Vogt \etal~1997). 
However, it is useful to consider the potentially critical issue of color
selection criteria.  The Bershady (1997) low redshift ($0.05 \leq z \leq 0.35$)
sample shows an offset of less than 0.5 mag, while the Rix \etal\ (1997)
spatially unresolved data at redshift $z \sim 0.25$ show evidence for a
magnitude offset of $\sim$1.5 relative to a local {\it blue} sample (\ie the
effect would be even greater if the data were compared to a general local
sample). Simard \& Pritchet (1996) chose the strongest \fOII\ emitters from
among a sample of emission-line galaxies (Simard 1996), and they find the
highest offset of all (2.5 $\pm$ 0.5 mag for a redshift of $z \sim 0.35$; note
the large scatter).  This suggests the bluest, most actively star forming
galaxies may show the largest offsets.  Forbes \etal\ (1996), whose sample was
not color selected, noted some correlation between their offsets and galaxy
colors in the sense that the galaxies with the largest offsets tended to be
blue.  Rix \etal\ (1997) find the same trend within their blue sample.  As
further example, Figure~\ref{TF} includes the two galaxies from Vogt
\etal~(1993) with observed optical rotation curves; one (SA68$-$2545.3) was
chosen specifically for its unusually strong \fOII\ flux, and this galaxy shows
a very large offset from the TF relation.  Taken together, this suggests that
for redshifts $z \gtrsim 0.2$, color may prove to be a good indicator of
luminosity evolution in field galaxies, distinguishing between an average,
stable population and a bluer, star-forming population with enhanced luminosity.

\subsection{Surface Brightness Evolution}

Changes in surface brightness levels in disk galaxies can provide an
independent determination of luminosity evolution --- provided scale lengths do
not evolve strongly --- and is particularly useful since it is independent of
$q_0$.  Freeman (1970) showed that disks in local spiral galaxies have a
near-uniform central surface brightness ($\mu_B$ = 21.65 mag arcsec$^{-2}$). 
Recently de Jong (1995) found a morphological-type dependence to the surface
brightness, and studied the effect of internal extinction, determining a value
of $\mu_B$ = 21.45 $\pm$ 0.76 mag arcsec$^{-2}$ for the case of spirals of
T-types 1--6 (RC3) with semitransparent disks.  Among high redshift ($z \sim
0.5$) galaxies, Forbes \etal\ (1996) concluded that the surface brightness
increases by 0.6 $\pm$ 0.1 mag with respect to local galaxies of similar mass. 
This is also in agreement with Colless \etal\ (1994).  Schade \etal\ (1996b,a)
find increases of $\sim$0.9 mag out to a redshift $z \sim 0.5$ and $1.6 \pm
0.1$ mag for disk galaxies at redshifts $0.5 < z < 1.1$, respectively (with no
correction for internal extinction).  For our sample, we compare disk sizes and
luminosities for the 15 disk galaxies, as plotted in Figure~\ref{sb} (we
exclude the ``double nucleus'' galaxy from \vfp, as its structure is complex). 
The single early-type spiral (NE4-1269-1248) appears to be a ``ring galaxy''
whose profile is difficult to fit. Scale length measurements for it range from
{0\farcs 4} to {1\farcs 0}; we adopt {0\farcs 6} (with large uncertainties) as
it appears most credible.  This galaxy also has a significantly larger B/D
ratio ($\sim$0.9) than that of the others ($\sim$0.1).  As is normal practice,
the disk scale lengths have not been corrected to face-on values.  Sources with
inclination $i \gtrsim 80$\deg\ may be systematically in error, due to
distortion of the surface brightness profiles from non-uniform extinction at
different radii (\cf Giovanelli \etal\ 1994).  Comparison is made with de
Jong's local galaxy sample, and distant galaxy measurements from Schade \etal\
(1996a).  We find our sample to have an overall offset of $0.59 \pm 0.13$ mag
with respect to local galaxies, in fairly good agreement with the offset in the
TF relation.  Though the highest redshift data ($z > 0.75$) generally lie
within the locus of the data points from Schade \etal\, the median offset in
our data is significantly less.  The apparent brightening at the high mass end
(seen also in the data of Forbes \etal~1996), could be caused by a bias toward
higher-than-average luminosities among the most distant objects, as well as by
luminosity evolution.

\section{Conclusions}

In summary, we have compared a set of 16 high redshift galaxies with a local TF
relation and find a modest amount of luminosity evolution ($\Delta$M$_B
\lesssim$ 0.4).  This conclusion is supported by an examination of the surface
brightness characteristics of the sample, which show evidence for evolution at
the level of $\sim$ 0.6 magnitude.  We find no evidence for deviation from a
linear TF relation at lower luminosities, down to a magnitude M$_B \sim -$19. 
The bluest galaxies within the sample have a slightly larger offset from the
local TF relation, which suggests (when taken in conjunction with results of
other studies) that the derived degree of luminosity evolution may depend
strongly on sample selection.

\acknowledgments

DEEP was established through the Center for Particle Astrophysics.  Funding was
provided by NSF grants AST-9529098, AST-922540, and AST-9120005;  NASA grants
AR-06337.08-94A, AR-06337.21-94A, GO-05994.01-94A, AR-5801.01-94A, and
AR-6402.01-95A.  JG acknowledges partial support from Spanish MEC grants
PB89--124 and PB93--456 and a UCM del Amo foundation fellowship; CG funding
from an NSF Graduate Fellowship; and JDL support from NASA grant HF-1048.01-93A.

\vfill\eject

\pagebreak

\begin{figure} 
	\caption{ ({\sc Plate})
HDF galaxies and their velocity curves.  Scales are identical and spatial axes
aligned for each set of three plots.  The WFPC2 $I$-band image of each galaxy
are shown ({\it top}), along with its \I, \V --\I, and M$_B$ magnitudes and the
\LRIS\ slit width and orientation.  Below is a $\sim$40\AA\ section of the
\LRIS\ spectrum centered around the redshifted \fOII\wave{3727} doublet ({\it
middle}).  In the velocity plots ({\it bottom}), points represent the observed
velocities: \fOII\ ($\bullet$), \Hbeta\ (filled $\triangle$), \fOIII\wave{5007}
($\circ$), and \Halpha\ (unfilled $\triangle$).  Error bars are internal errors
from the line-fitting technique.  Data are shown only where the \SNr\ was
adequate; the \Hbeta\ and \fOIII\ lines are not fit for several sources because
of weak line strength, interference caused by night-sky OH emission lines, or
fringing in the red portion of the spectrum.  The solid line is the fit to the
model emission line(s), and the intrinsic circular velocity of the model is
noted (parentheses indicate lower-quality sources).  A small tick marks the
measured disk scale length.  {\it Individual notes:} (IE4-1304-1007) all
emission lines plus \fNII\wave{6583} were traced (only three are shown for
legibility), but the strong bar and face-on nature of this galaxy make this a
poor candidate; (IE2-0201-0520) measured \fOII\ velocities are somewhat larger
than \fOIII\ and \Hbeta, so the median velocity at each radius was fit;
(NE4-1269-1248) early-type ring galaxy with \fOII\ flux concentrated in core,
\fOIII\ and \Hbeta\ too weak to trace; (OE4-0981-1325) \fOIII, \Hbeta\ too weak
to trace; (SE2-0304-0307) note good agreement between \fOII\ and both \fOIII\
and \Hbeta, observed with different gratings; (NW2-0358-0561) note excellent
agreement of \Hbeta\ and \fOII; (IE2-0229-0212) night sky line bisects \fOII\
emission, leading to large uncertainties.  }
	\label{PLATE} 
\end{figure}

\begin{figure}
	\caption{ 
High-redshift Tully-Fisher diagram, plotting \Vmax\ vs $B$ luminosity.  We show
our data (numbered in increasing redshift) and those from Vogt \etal\ (1993;
1996) compared to the relationship (solid line) based on \HI\ velocity width
measurements of a restricted set of 32 local cluster spirals (Pierce \& Tully
1992).  Dashed lines are the 3$\sigma$ limits to the dispersion in this local
relation.  Velocities have been corrected by sin\thinspace$i$ and the
magnitudes corrected for internal extinction.  We adopt \Hconst{75} and $q_0 =
0.05$.  Assuming the same slope as that of the local sample, the weighted fit
(dotted line) to our 11 high-quality points (filled symbols) is offset from the
local relation by $0.36 \pm 0.13$ mag toward higher luminosity.  }
	\label{TF}
\end{figure}

\begin{figure}
	\caption{ 
Disk surface brightness evolution diagram.  Exponential disk scale lengths,
$R_d$ (uncorrected for inclination) are plotted versus disk $B$ luminosity for
the 15 disk galaxies in our sample (new sources are numbered in increasing
redshift as in Figure~\ref{TF}).  The five sources with redshift $0.75 \leq z
\leq 1$ are circled.  Open symbols denote disks with apparent inclination $i
\geq 80$\deg, which may suffer from large systematic errors (see text).  Note
that $R_d$ was measured from the major-axis profile, with a disk and
$r^{1/4}$-law bulge fit simultaneously.  The median bulge-to-disk ratio is 0.1;
this value is generally consistent with the Forbes \etal\ (1996) and Colless
\etal\ (1994) sample of faint field galaxies.  The region occupied by local
spirals (from de Jong 1995) is lightly shaded; that populated by the galaxies
at redshift $\bar{z} \sim 0.7$ from Schade \etal\ (1996a) is dark (arrows point
to median values).  The solid line is the offset ($0.59 \pm 0.13$ mag) found by
a weighted fit to the 12 disks with inclination angles less than 80\deg, while
the dashed line indicates the luminosity offset ($0.36 \pm 0.13$ mag) found
with respect to the local TF relation (Figure~\protect\ref{TF}). }
	\label{sb}
\end{figure}
 
\end{document}